\documentclass[a4paper]{article}
\usepackage{spconf,amsmath,graphicx,epsfig, bm}
\usepackage{color}
\usepackage[table]{xcolor}
\usepackage{multirow}
\usepackage[table]{xcolor}
\usepackage{float}
\usepackage{textcomp}
\usepackage{caption}
\usepackage{subcaption}
\usepackage{verbatim}
\usepackage{soul}

\title{Data Generation Using Pass-phrase-dependent Deep Auto-encoders for Text-Dependent Speaker Verification}

\name{Achintya Kumar Sarkar$^1$, Md Sahidullah$^2$, Zheng-Hua Tan$^3$}
\address{$^1$Indian Institute of Information Technology, Sri City, India\\
$^2$ Universit\'{e} de Lorraine, CNRS, Inria, LORIA, F-54000, Nancy, France\\
$^3$Department of Electronic Systems, Aalborg University, Denmark\\
{\small \tt sarkar.achintya@gmail.com, md.sahidullah@inria.fr, zt@es.aau.dk}}
\begin{document}
\ninept
\maketitle

\begin{abstract}

In this paper, we propose a novel method that trains pass-phrase specific deep neural network (PP-DNN) based auto-encoders for creating augmented data for text-dependent speaker verification (TD-SV). Each PP-DNN auto-encoder is trained using the utterances of a particular pass-phrase available in the target enrollment set with two methods: (i)~ transfer learning and (ii)~ training from scratch. Next, feature vectors of a given utterance are fed to the PP-DNNs and the output from each PP-DNN at frame-level is considered one new set of generated data. The generated data from each PP-DNN is then used for building a TD-SV system in contrast to the conventional method that considers only the evaluation data available. The proposed approach can be considered as the transformation of data to the pass-phrase specific space using a non-linear transformation learned by each PP-DNN. 
The method develops several TD-SV systems with the number equal to the number of PP-DNNs separately trained for each pass-phrases for the evaluation. 
Finally, the scores of the different TD-SV systems are fused for decision making. Experiments are conducted on the RedDots challenge 2016 database for TD-SV using short utterances. Results show that the proposed method improves the performance for both conventional cepstral feature and deep bottleneck feature using both Gaussian mixture model- universal background model (GMM-UBM) and i-vector framework.

\end{abstract}


\section{Introduction}
Speaker verification (SV) aims to verify a person based on their voice signal. This is realized by using either \emph{text-independent} (TI) or \emph{text-dependent} (TD) mode. The speakers in TI-SV systems have the flexibility to speak any sentence or text during both the enrollment and test phases. 
Whereas in TD-SV, speakers are constrained to speak the predefined pass-phrases during both enrollment and test. Since TD-SV maintains the matched phonetic condition between the enrollment and test phases, it gives low error rates in SV using short utterances. This makes TD attractive for real-world applications. 

In the literature, many techniques have been proposed for the improvement of TD-SV. Model domain methods include Gaussian mixture model- universal background model (GMM-UBM) \cite{reynold00}, i-vector or total variability modeling \cite{Deka_ieee2011} and x-vector \cite{conf/icassp/SnyderGSPK18}, and in the feature domain cepstral Mel-frequency cepstral coefficients (MFCC) and bottleneck (BN) \cite{DBLP:journals/taslp/SarkarTTSG19,achintya_axriv2020} feature-based techniques are commonly used. Though the x-vector systems give promising results in TI-SV, they are not successful so far in TD-SV possibly due to the limited training data~\cite{zeinali2019short}. All those techniques require a large amount of audio-data for training its speaker-independent (SI) model parameters. Generally, the SI hyper-parameters in those modeling techniques, (e.g., GMM-UBM or total variability space) are trained using the data/pass-phrase sets which are different from the evaluation set. 
This is done due to the lack/unavailability of a large amount of data (pass-phrases) matched to the evaluation set and is an open problem to the TD-SV research communities.

Recently, several data augmentation techniques have been proposed in the literature for creating additional data under low resource applications: 
vocal tract length perturbation \cite{Jaitly_vocaltract}, SpecAugment (deformation of log Mel spectrogram with frequency masking) \cite{Park_2019},  random image warping \cite{24792} on image processing,  mixing noise or other speech files with the given raw speech signal \cite{noise_mix,Lasseck2018}, and applying impulse (IR) response (of hall room, classroom) on the given raw speech signal \cite{Stewart2010}.
The effectiveness of data augmentation has been proven in various studies including speech recognition \cite{Park_2019}, speaker recognition \cite{conf/icassp/SnyderGSPK18,achintya_axriv2020}  and image processing \cite{Shorten2019}. 

In the context of  speaker verification, augmented data in form of added noise with existing training data are conventionally used for training the SI  model parameters in \emph{TI-SV}, e.g. GMM-UBM \cite{I4U2013, michelsanti2017conditional}, DNNs \cite{conf/icassp/SnyderGSPK18}, total variability space in i-vector \cite{conf/icassp/SnyderGSPK18}, and in post-processing/scoring step, e.g., probabilistic linear discriminate analysis (PLDA) \cite{conf/icassp/SnyderGSPK18,achintya2012}. \emph{However, none of these deals with data augmentation for speaker enrollment and test phases.} A limited number of studies are made where the noisy version of training speech utterances/speaker enrollment data has been included in the enrollment phase for building a noise-robust model for spoofing detection \cite{7528399} and TI-SV recognition \cite{I4U2013}. These works mostly use multi-conditional training, which is a classic approach for improving noise robustness. \emph{However, to the best of our knowledge, there is no study on generating auxiliary data for enrollment and verification in TD-SV.}

\begin{figure*}[t]
\centering\includegraphics[height=2.7cm,width=17.0cm]{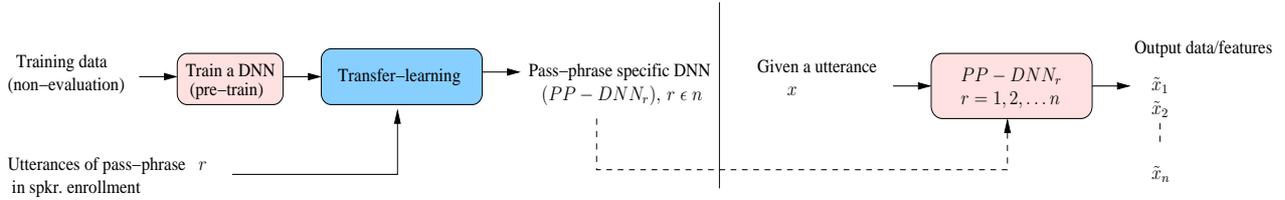}
\caption{Illustration of data generation using pass-phrase specific DNN with transfer learning for TD-SV.}
\label{fig:fig1}
\end{figure*}

This motivates us to investigate deep neural network (DNN) for generating additional data for TD-SV. First, we train pass-phrase specific autoencoder referred to here as \emph{pass-phrase specific DNN} (PP-DNN) using the utterances of the particular pass-phrase. Then audio-data for enrollment and test are processed with the PP-DNNs and the output of PP-DNNs are used as auxiliary data for developing TD-SD system.
The proposed method can be viewed as the process of mapping a speech utterance onto the evaluation pass-phrase-specific DNN space using non-linear transformation. This allows generating several copies of the data-sets by processing the utterances with PP-DNNs.  
The transformation of evaluation pass-phrase through different PP-DNNs can be considered as the capturing of cross-pass-phrase information relevant for the TD-SV. The generated data with the proposed method is then used to build a TD-SV \emph{per evaluation pass-phrase, i.e. PP-DNN} with the standard TD-SV methods such as GMM-UBM or i-vector. Finally, the scores of the sub-systems based on different PP-DNN generated data are fused for decision making.    

For training the PP-DNNs, two techniques are considered: (i)~PP-DNNs are derived from a pre-trained DNN with transfer learning \cite{Dong2015,Yim_2017_CVPR,7472128,7953073,9054468}. 
In the second method, PP-DNNs are trained \emph{from scratch}. 
We demonstrate that the proposed method outperforms the conventional systems based on cepstral and BN features with GMM-UBM and i-vector techniques. 

The paper is organized as follows: Sec. \ref{Sec:TV-technique} describes the TD-SV techniques. The proposed method and experimental setup are presented in Secs. \ref{sec:prop} \&  \ref{sec:exp}, respectively.  Results and discussions are described in Sec.\ref{sec:resd}. Finally, the paper is concluded in Sec.\ref{sec:con}.

\section{Text-dependent speaker verification}
\label{Sec:TV-technique}
The TD-SV system uses a frame-level acoustic feature followed by a speaker modeling technique. In this work, we use MFCC. In addition, we also extract frame-level bottleneck (BN) features. For extracting the BN, a \emph{DNN} is trained which discriminates the speakers at the output layers with cross-entropy based objective function,

\begin{equation}
L (\theta) = - \frac{1}{N} \sum_{t=1}^{N} y_t\log p(\bm{x}_t, \theta)
\end{equation}

where $L$, $\theta$, $y_t$, $\bm{x}_t$ and $p(.)$ denote the loss, parameters of DNN, the class label of the $t$-th input feature vector and speaker posterior at the DNN output layer, respectively.
Next, the output from a particular hidden layer of the DNN is projected onto a low dimensional space \cite{Yuan2015} to extract the BN. Next, we use the GMM-UBM and i-vector methods for speaker modeling and scoring.
\subsection{GMM-UBM}
\label{sec:GMM_tech}
In the GMM-UBM method, speaker dependent models are derived from a GMM-UBM using the enrollment data for the particular speaker with \emph{maximum-a-posteriori} (MAP) adaptation~\cite{reynold00}. During test, the feature vectors of the test utterance $\mathbf{X}=\left\{\mathbf{x_1},\mathbf{x_2},\ldots, \mathbf{x_k} \right\}$ is scored against the claimant $\lambda_r$ (obtained in the enrollment phase) and GMM-UBM $\lambda_{ubm}$ models, respectively. Finally, the log-likelihood ratio, $\Lambda(\bm{X}) =\frac{1}{k} \left[ \log p(\bm{X}|\lambda_r)- \log p(\bm{X}|\lambda_{ubm}) \right]$ is calculated using scores between the claimant and UBM models for decision making.

\subsection{i-vector}
\label{sec:i-vect}
In the i-vector method, the i-vector of a given speech utterance for a speaker is obtained by decomposing the speaker and channel-dependent GMM super-vector $\bm{M}$ as $\bm{M} = \bm{m} + \bm{Tw}$ where $\bm{m}$ denotes the speaker-independent GMM super-vector and {$\bm{w}$ is called an i-vector}~\cite{Deka_ieee2011}. Here $\bm{T}$ is the total variability space on a subspace of $\bm{m}$, where speaker and channel information is assumed to be dense. During the training phase, each target is represented by an average i-vector computed over all of its enrollment i-vectors. During the test, the i-vector of the test utterance is scored against the claimant-specific i-vector by PLDA~\cite{SenoussaouiInterspch2011}.

\section{Proposed method for data generation}
\label{sec:prop}

In the conventional/baseline TD-SV system, the speakers are enrolled with speech phrases from multiple sessions available in the evaluation set as per the evaluation plan. Here we propose a method for phrase-specific transformation using PP-DNN where features of a speech utterance are transformed at frame-level for generating multiple sets for training systems in an identical manner.



\subsection{Pass-phrase-dependent DNN auto-encoder} 
At the first step, the PP-DNNs are trained using often limited utterances  of particular pass-phrases used for the evaluation. The objective is to reconstruct the given input at the output layer and the objective function to be minimized is the reconstruction loss (e.g., mean squared error) plus $l2$ regularization \cite{8700495}. It can be defined as,

\begin{equation}
loss = \frac{1}{N} \sum_{t=1}^{N} \parallel \bm{x}_t - \bm{\hat{x}}_t \parallel^2 + \;l_2 
\end{equation}

\noindent where $\bm{x}_t$, $\bm{\hat{x}}_t$  denote, respectively, the $t$-th input frame and corresponding reconstructed input at the output layer of the DNN.  
Two strategies are considered for training PP-DNNs: {\bf (i) Transfer learning:}  PP-DNNs are derived from a pre-trained DNN  with transfer learning concept {(ii) \bf Learning from scratch:} PP-DNNs are trained from \emph{scratch} with limited audio-data for specific pass-phrases. The two approaches will reflect the impact of an apriori knowledge on PP-DNNs modeling and so on the performance of TD-SV \emph{specially when a very limited amount of pass-phrase-wise data is available for training the systems}.

\subsection{Creation of new data} 
In the second step, the feature vectors for a given utterance is fed to the PP-DNNs and the frame-wise output from \emph{each PP-DNN} is considered as the generated new features. This transformation is applied to the entire data set, including training, development and evaluation (both enrolment and test) data. The proposed system is illustrated in Fig.~\ref{fig:fig1}. Based on the number of PP-DNNs (say $n$, i.e., the number of pass-phrases in the evaluation database), the proposed approach generates $n$ sets of different features and they are separately used for the system development, i.e., for training GMM-UBM, T-space, and PLDA. It can be expressed as,

\begin{equation}
\bm{\hat{x}}_i = f_{pp\-dnn}^{i}(\bm{x}), \quad   i=1, \ldots, n \label{eq:trans}
\end{equation}

\noindent where $\bm{\hat{x}}_i$ denotes the generated feature vectors using the $i$-th PP-DNN from a utterance $\bm{x}$. A TD-SV system is built using the generated data (feature vectors) from a particular PP-DNN. 
Finally, the scores of the PP-DNN systems are averaged with equal importance,
    
    \begin{equation}
        {f_{score}} = \frac{1}{n}\sum_{j=1}^{n} Sys_{score}^j \label{eq:fusion}
    \end{equation}
    
We consider the MFCC and BN features and hence we develop two feature-based systems called \emph{PP-DNN MFCC and PP-DNN BN}, respectively. Note that PP-DNN MFCC features are used for extracting the PP-DNN BN.




\section{Experimental setup}
\label{sec:exp}
Experiments are performed on male speakers in RedDots challenge 2016 database (task \emph{m-part-01}) as per protocols in \cite{RedDots}. 
There are three enrollment sessions to train the particular pass-phrase-wise target speaker model. The utterances are of very short duration on an average of 2-3s per speech signal and recorded over $10$ pass-phrases. A disjoint set of nine speakers' data (approximately $148$ files per pass-phrase, excluded from the evaluation) are considered as a development set. The remaining speakers are considered for the evaluation \cite{Kinnunen2016}. This gives $248$ target models. The pass-phrases available in the speaker enrollment and development sets are used for the PP-DNN training ($\approx$ $223$ speech files per model) and give total of $10$ PP-DNNs. Table \ref{table:trial_info} shows the number of different trials available for the system evaluation.

\begin{table}[h]
\caption{\it Number of trials for system evaluation.}
\vspace*{-0.3cm}
\begin{center}
\begin{tabular}{|l|l|l|l|}\cline{1-4}
\# of & \multicolumn{3}{c|}{\# of trials in non-target type} \\ 
Genuine      & Target  & Impostor & Impostor \\
trials     &-wrong (TW)  &-correct (IC) & -wrong (IW)\\ \hline 
2119        & 19071        & 62008            & 557882  \\  \hline
\end{tabular}
\end{center}
\label{table:trial_info}
\end{table}

\begin{table*}[ht]
\begin{center}
\caption{Performance of TD-SV using different PP-DNNs and different numbers of hidden layers on the RedDots database (m-part-01 task).}
\setlength\tabcolsep{6pt}
\footnotesize{
\begin{tabular}{|l|l|l|l|l|l||l|l|}\cline{1-8}
System/\# hidden  &\multicolumn{3}{c|}{Non-target type [\%EER/(MinDCF$\times$ 100)]}     & \multicolumn{2}{c|}{Average (EER/MinDCF)}  & \multicolumn{2}{c|}{Average (EER/MinDCF)}         \\
layer (PP-DNNs)          & Target-wrong            &Impostor-correct   & Impostor-wrong     &    & +baseline  & {\bf PP-DNNs Scratch}  & + baseline\\ \hline
Baseline         & 3.91/1.49    &	2.67/1.23  & 0.89/0.24  & 2.49/0.99      & -          & - & - \\ 
   & \multicolumn{4}{c|}{\multirow{1}{*}{{{\bf Transfer-learning}}}} & & &   \\      
PP-DNNs / 1              & 3.49/1.42       &	{\bf 2.50}/1.17  & {\bf 0.84}/0.18  & {\bf 2.28}/0.92 & 2.28/0.92      & 2.34/0.97        & 2.29/0.93     \\
\hspace*{+1.2cm}/2       & 3.58/1.41       &	2.59/1.18        & 0.94/0.19        & 2.37/0.93       & 2.37/0.92      & 2.36/0.94        & 2.29/0.93\\
\hspace*{+1.2cm}/3       &{\bf 3.48}/1.43 &2.59/1.20             &{\bf 0.84}/0.17   & 2.31/0.93       &{\bf 2.26}/0.92 & {\bf 2.27}/0.93  & {\bf 2.24}/0.93 \\
\hspace*{+1.2cm}/4       & 3.63/1.44      &{\bf 2.50}/1.17       & 0.89/0.20	    & 2.34/0.94       & 2.31/0.93     & 2.29/0.93        & 2.35/0.93\\
\hspace*{+1.2cm}/5       & 3.56/1.44	  & {\bf 2.50}/1.18      & 0.89/0.19        & 2.31/0.94       & 2.31/0.92      & 2.29/0.92        & 2.29/0.96\\ \hline 
\end{tabular} \\
}
\label{table:analysis}
\end{center}
\end{table*}

$57$-dimensional MFCC feature vectors ($19$ static and their $ \Delta, \Delta\Delta$) are extracted from speech signals using a $20 ms$ Hamming window and a $10ms$ frame-shift with RASTA filtering \cite{Hermanksy94}. An open-source robust voice activity detection (rVAD) \cite{TanSD20} is applied to discard the speech frames with lower energies. The selected frames are processed with utterance-level cepstral mean and variance normalization (CMVN).
A GMM-UBM with $512$ mixture components and diagonal co-variance matrices is trained using $6300$ speech files from the TIMIT database consisting of $630$ speakers. The same data set is also used for training PCA projection matrix during BN features extraction. During MAP adaptation, we consider three iterations and relevance factor of $10$.

For T-space, pre-trained DNN, PLDA, and DNN for the BN feature extraction, $72764$ utterances (out of which $1000$ utterances are left out for validation) over $27$ pass-phrases (excluding the pass-phrases common in RedDots database) from the RSR2015 database \cite{RSR2015} consisting of $157$ male and $143$ female speakers are used. 
The PP-DNN auto-encoder consists of hidden layers with $512$ neurons, ReLU activation function and one linear output layer \cite{chung2020generative}. We experiment with various numbers of hidden layers. 
The learning rate, dropout, batch size, and the number of the epoch are considered, respectively $0.001, 0.01, 1024$, and $50$. In transfer learning and from scratch cases, $30$ epochs are followed as the validation loss was decaying very slowly (as very limited data is available for training PP-DNNs). 

For the BN extraction, DNN consists of a seven-layer feed-forward network, $1024$ neurons per hidden layer, sigmoid activation function and 300 nodes at the output (number of speakers in the DNN training dataset). The input layer considers the context window of $11$ frames (i.e. $5$ frames left, current frame, $5$ frames right). Each hidden layer consists of $1024$ neurons. The frame-level output from the fourth hidden layer of DNNs for BNs is projected using PCA onto $57$ dimensional space. The dimension is set to 57 for a fair comparison with MFCC as per \cite{DBLP:journals/taslp/SarkarTTSG19}. We use TensorFlow toolkit for training the DNNs~\cite{tensorflow2015-whitepaper}.

The i-vector ($400$ dimensional) system and PLDA scoring (with default parameters) are developed using the Kaldi toolkit\cite{Povey_ASRU2011}. In PLDA, the utterances of the same pass-phrase from a particular speaker are treated as an individual class and this gives $8100$ classes (4239 males and 3861 females) in PLDA. System performance is measured in terms of equal error rate (EER) and minimum detection cost function (minDCF) as per the 2008 SRE \cite{SRE08}.

\section{Results and discussion}
\label{sec:resd}

In this section, we first analyze the performance of TD-SV with the data generated by the PP-DNNs for two different training methods and the different number of hidden layers. The results are shown in Table \ref{table:analysis} on part 1 of RedDots database. We observe that the performance of TD-SV (in terms of average EER) for different PP-DNNs based data is very close irrespective of the number of layers in their DNNs and training methodology. The fusion of the PP-DNNs system with baseline gives a marginal   reduction in the error rate.

\begin{figure}[h]
\centering\includegraphics[height=6.0cm,width=9.5cm]{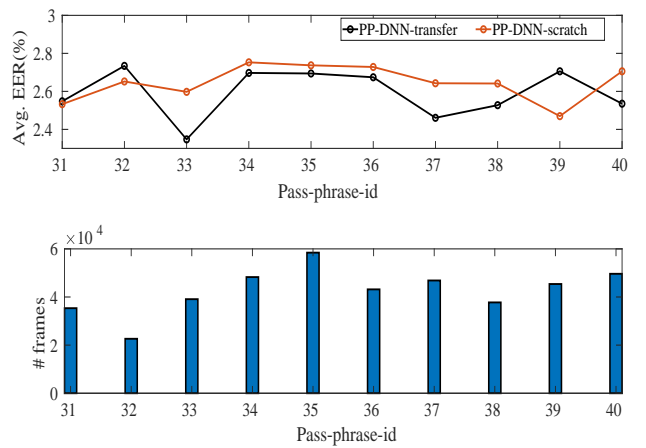}
\caption{Upper penal: performance comparison of TD-SV with data created using different PP-DNNs (for 3 Hidden layers) methods; lower penal: number of frames available per pass-phrase during training. } 
\label{fig:fig2}
\end{figure} 

To look at the behaviors of the PP-DNNs trained with different methods, we compare the performance of TD-SV for each PP-DNN  (3 hidden layers) generated data in Fig. \ref{fig:fig2}. From Fig.\ref{fig:fig2}, it can be noticed that the performances of the TD-SV systems are very different for each PP-DNN based data/feature.
However, the average EER values of the systems (after score fusion using Eq.(\ref{eq:fusion})) are very close as in Table \ref{table:analysis}. This indicates that both the  transfer-learning and scratch method based PP-DNNs  are able to capture similar speaker relevant information for the TD-SV at the end.
For simplicity,  in the rest of the paper, we focus on the TD-SV with transfer-learning based PP-DNN data in further analysis and for comparison with baseline. 

\begin{figure}[h]
\centering\includegraphics[height=5.0cm,width=9.8cm]{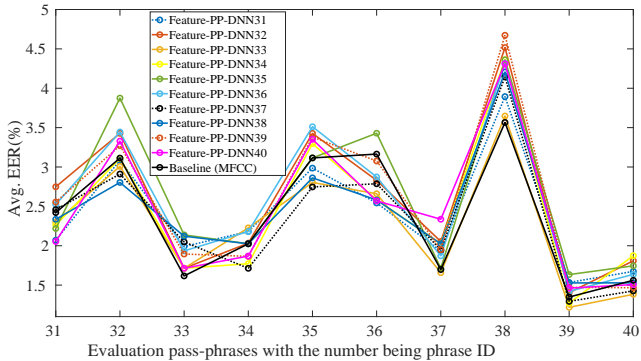}
\caption{Comparison of the performance of TD-SV across pass-phrases for different PP-DNN based systems and the baseline.} 
\label{fig:fig3}
\end{figure}

To further investigate the behaviours of PP-DNN based systems, we compare the performance of TD-SV for each pass-phrase in the evaluation set using the PP-DNN (with transfer-learning) based systems and the baseline system as shown in Fig.\ref{fig:fig3}. 
 It can be observed from Fig.\ref{fig:fig3}, that the performance of TD-SV (in terms of average EER) quite differs across the systems.  This shows that each PP-DNN generated data impact differently on the TD-SV, i.e., PP-DNNs are different though a very small amount of data is available for their training. 
Moreover, the performance of the PP-DNN feature-based systems is similar to that of the baseline MFCC and at the same time shows some variation, which is a positive observation as it indicates the different characteristics of these systems and the potential to combine them to improve the overall performance as already shown in Table \ref{table:analysis}. 
 

\subsection{Impact of bottleneck features}
Table \ref{table:BN} compares the performance of TD-SV for the proposed method with  baseline using MFCC and BN features on the RedDots database. 
Similarly, for the proposed PP-DNN framework, the relative improvement with BN feature is considerably higher. 
 
\begin{table}[h]
\begin{center}
\caption{Comparison of performance of TD-SV using the proposed method with the baseline for different features on RedDots (m-part-01 task) with the GMM-UBM technique.}
\setlength\tabcolsep{6pt}
\footnotesize{
\begin{tabular}{|l|l|l|l|l|}\cline{1-5}
System      &\multicolumn{3}{c|}{Non-targets [\%EER/(minDCF$\times$ 100)]} & Avg. (EER/           \\
-Feature                 & TW           &IC    & IW  & minDCF)     \\ \hline
{\bf Baseline}                &              &      &     & \\
-MFCC          & 3.91/1.49    &	2.67/1.23  & 0.89/0.24   & 2.49/0.99 \\
-BN            & 3.44/1.46    & 2.90/1.32  & 0.99/0.30   & 2.44/1.02\\ 
 \hline
{\bf PP-DNNs}         &              &      &     & \\
-MFCC            & 3.48/1.43    & 2.59/1.20     & 0.84/0.17 & 2.31/0.93    \\
-BN     &  {\bf 2.99/1.21}   & {\bf 2.36/1.10}     & {\bf 0.51/0.16}  &{\bf 1.96/0.83}    \\
 \hline
\end{tabular}
}
\label{table:BN}
\end{center}
\end{table}

\subsection{Performance with the i-vector-PLDA technique}
Table \ref{table:ivector} compares the performance of TD-SV for proposed method with baseline using i-vector-PLDA technique on the RedDot (m-part-01 task). We observe that the proposed method shows a more relative reduction in EER compared to the improvement obtained with GMM-UBM system in the previous subsection.

\begin{table}[h]
\begin{center}
\caption{Comparison of performance of TD-SV using the proposed method with  baseline on RedDots (m-part-01 task) with i-vector.}
\setlength\tabcolsep{6pt}
\footnotesize{
\begin{tabular}{|l|l|l|l|l|}\cline{1-5}
System      &\multicolumn{3}{c|}{Non-targets [\%EER/(MinDCF$\times$ 100)]} & Avg.(EER/           \\
-Feature         & TW           &IC    & IW     & /minDCF)  \\ \hline
Baseline        &              &               &       & \\
-MFCC           & 5.98/2.56    &4.38/1.84	 & 1.32/0.45  & 3.89/1.62     \\
-BN             & 5.88/2.72	   & 4.43/2.05	 & 1.42/0.53  &	3.91/1.76\\
 \hline
PP-DNNs         &    &         &      & \\
-MFCC           & 4.48/2.06    & 3.30/1.29   & 0.79/0.29   & {\bf 2.98/1.21}  \\
-BN             & 4.24/1.19    & 3.39/1.32   &  0.75/0.23  & {\bf 2.79/1.15}   \\
 \hline
\end{tabular}
}
\label{table:ivector}
\end{center}
\end{table}

\section{Conclusion}
\label{sec:con}
In this paper, we introduced PP-DNN based auto-encoders for creating additional data for TD-SV, where PP-DNNs are trained using utterance of pass-phrases used for the evaluation of TD-SV. The utterances were processed at frame-level by the PP-DNNs and the output from each PP-DNN is considered as the set of generated new data for TD-SV. We also studied the impact of transfer learning while training the PP-DNNS. Our TD-SV experiments on RedDots corpus with generated data demonstrate consistent improvement over original data. Our method is simple but effective in reducing the EERs and cost metrics. The work can be extended by adopting advanced DNN architectures such as recurrent and variational for improving the data generation process. Advanced fusion strategies can be explored in place of simple score fusion adopted in our work.


\clearpage
\ninept
\bibliographystyle{IEEEbib}
\bibliography{strings,References}

\begin{thebibliography}{10}

\bibitem{reynold00}
D.~A. Reynolds, T.~F. Quatieri, and R.~B. Dunn,
\newblock ``{S}peaker verification using adapted gaussian mixture models,''
\newblock {\em Digital Signal Processing}, vol. 10, pp. 19--41, 2000.

\bibitem{Deka_ieee2011}
N.~Dehak, P.~Kenny, R.~Dehak, P.~Ouellet, and P.~Dumouchel,
\newblock ``{F}ront-end factor analysis for speaker verification,''
\newblock {\em IEEE Trans. on Audio, Speech and Language Processing}, vol. 19,
  pp. 788--798, 2011.

\bibitem{conf/icassp/SnyderGSPK18}
D.~Snyder, D.~Garcia{-}Romero, G.~Sell, D.~Povey, and S.~Khudanpur,
\newblock ``{X}-vectors: Robust dnn embeddings for speaker recognition,''
\newblock in {\em Proc. of ICASSP}, 2018, pp. 5329--5333.

\bibitem{DBLP:journals/taslp/SarkarTTSG19}
A.~K. Sarkar, Z.-H. Tan, H.~Tang, S.~Shon, and J.~R. Glass,
\newblock ``{T}ime-contrastive learning based deep bottleneck features for
  text-dependent speaker verification,''
\newblock {\em {IEEE/ACM} Trans. Audio, Speech {\&} Language Processing}, vol.
  27, no. 8, pp. 1267--1279, 2019.

\bibitem{achintya_axriv2020}
A.~K. Sarkar, H.~Sarma, and Z.-H.~Tan P.~Dwivedi,
\newblock ``{D}ata augmentation enhanced speaker enrollment for text-dependent
  speaker verification,''
\newblock in {\em arXiv:2007.08004}, 2020.

\bibitem{zeinali2019short}
Hossein Zeinali, Kong~Aik Lee, Jahangir Alam, and Lukas Burget,
\newblock ``Short-duration speaker verification (sdsv) challenge 2020: the
  challenge evaluation plan,''
\newblock {\em arXiv preprint arXiv:1912.06311}, 2019.

\bibitem{Jaitly_vocaltract}
N.~Jaitly and G.~E. Hinton,
\newblock ``{V}ocal tract length perturbation (vtlp) improves speech
  recognition,''
\newblock in {\em Proc. of ICML}, 2013.

\bibitem{Park_2019}
D.~S. Park et~al.,
\newblock ``{S}pecaugment: A simple data augmentation method for automatic
  speech recognition,''
\newblock pp. 2613--2617, 2019.

\bibitem{24792}
F.~L. {Bookstein},
\newblock ``{P}rincipal warps: Thin-plate splines and the decomposition of
  deformations,''
\newblock {\em IEEE Transactions on Pattern Analysis and Machine Intelligence},
  vol. 11, no. 6, pp. 567--585, 1989.

\bibitem{noise_mix}
A.~Hannun et~al.,
\newblock ``{D}eep speech: Scaling up end-to-end speech recognition,''
\newblock in {\em in arXiv, 2014}.

\bibitem{Lasseck2018}
M.~Lasseck,
\newblock ``{A}udio-based bird species identification with deep convolutional
  neural networks,''
\newblock in {\em In: Working Notes of CLEF 2018 (C{r}oss Lan{g}uage Evaluation
  Fo{r}um)}.

\bibitem{Stewart2010}
R.~Stewart and M.~Sandler,
\newblock ``{D}atabase of omnidirectional and b-format room impulse
  responses,''
\newblock in {\em Proc. of ICASSP}, 2010, pp. 165--168.

\bibitem{Shorten2019}
C.~Shorten and T.~M. Khoshgoftaar,
\newblock ``{A} survey on image data augmentation for deep learning,''
\newblock {\em Journal of Big Data}, 2019.

\bibitem{I4U2013}
R.~Saeidi et~al.,
\newblock ``{I4U} ssubmission to nist sre 2012: A large-scale collaborative
  effort for noise-robust speaker verification,''
\newblock in {\em Proc. of Interspeech}, 2013, pp. 1986--1990.

\bibitem{michelsanti2017conditional}
D.~Michelsanti and Z.-H. Tan,
\newblock ``{C}onditional generative adversarial networks for speech
  enhancement and noise-robust speaker verification,''
\newblock in {\em Proc. of Interspeech}, 2017, pp. 2008--2012.

\bibitem{achintya2012}
A.~K. Sarkar, D.~Matrouf, P.~M. Bousquet, and J.~F. Bonastre,
\newblock ``{S}tudy of the effect of i-vector modeling on short and mismatch
  utterance duration for speaker verification,''
\newblock in {\em Proc. of Interspeech}, 2012, pp. 2662--2665.

\bibitem{7528399}
H.~{Yu}, A.~{Sarkar}, D.~A.~L. {Thomsen}, Z.~{Tan}, Z.~{Ma}, and J.~{Guo},
\newblock ``{E}ffect of multi-condition training and speech enhancement methods
  on spoofing detection,''
\newblock in {\em Proc. of SPLINE}, 2016, pp. 1--5.

\bibitem{Dong2015}
D.~Wang and T.~F. Zheng,
\newblock ``{T}ransfer learning for speech and language processing,''
\newblock in {\em arXiv:1511.06066}.

\bibitem{Yim_2017_CVPR}
J.~Yim, D.~Joo, J.~Bae, and J.~Kim,
\newblock ``{A} gift from knowledge distillation: Fast optimization, network
  minimization and transfer learning,''
\newblock in {\em Proc. of CVPR}, 2017, pp. 7130--7138.

\bibitem{7472128}
H.~{Lim}, M.~J. {Kim}, and H.~{Kim},
\newblock ``Cross-acoustic transfer learning for sound event classification,''
\newblock in {\em Proc. of ICASSP}, 2016, pp. 2504--2508.

\bibitem{7953073}
J.~{Cui} et~al.,
\newblock ``{K}nowledge distillation across ensembles of multilingual models
  for low-resource languages,''
\newblock in {\em Proc. of ICASSP}, 2017, pp. 4825--4829.

\bibitem{9054468}
A.~{Abad}, P.~{Bell}, A.~{Carmantini}, and S.~{Renais},
\newblock ``Cross lingual transfer learning for zero-resource domain
  adaptation,''
\newblock in {\em Proc. of ICASSP}, 2020, pp. 6909--6913.

\bibitem{Yuan2015}
Y.~Liu, Y.~Qian, N.~Chen, T.~Fu, Y.~Zhang, and K.~Yu,
\newblock ``{D}eep {F}eature for {T}ext-dependent {S}peaker {V}erification,''
\newblock {\em Speech Communication}, vol. 73, pp. 1--13, 2015.

\bibitem{SenoussaouiInterspch2011}
M.~Senoussaoui et~al.,
\newblock ``{M}ixture of {PLDA} {M}odels {I}n {I}-{V}ector {S}pace {F}or
  {G}ender-{I}ndependent {S}peaker {R}ecognition,''
\newblock in {\em Proc. of Interspeech}, 2011, pp. 25--28.

\bibitem{8700495}
A.~{Rahangdale} and S.~{Raut},
\newblock ``{D}eep neural network regularization for feature selection in
  learning-to-rank,''
\newblock {\em IEEE Access}, vol. 7, pp. 53988--54006, 2019.

\bibitem{RedDots}
``The reddots challenge: Towards characterizing speakers from short
  utterances,''
  https://sites.google.com/site/thereddotsproject/reddots-challenge.

\bibitem{Kinnunen2016}
T.~Kinnunen et~al.,
\newblock ``{U}tterance verification for text-dependent speaker recognition: a
  comparative assessment using the reddots corpus,''
\newblock in {\em Proc. of Interspeech}, 2016, pp. 430--434.

\bibitem{Hermanksy94}
H.~Hermanksy and N.~Morgan,
\newblock ``{R}asta processing of speech,''
\newblock {\em IEEE Trans. on Speech and Audio Processing}, vol. 2, pp.
  578--589, 1994.

\bibitem{TanSD20}
Z.-H. Tan, A.~K. Sarkar, and N.~Dehak,
\newblock ``{rVAD}: An unsupervised segment-based robust voice activity
  detection method,''
\newblock {\em Computer Speech \& Language}, vol. 59, pp. 1--21, 2020.

\bibitem{RSR2015}
A.~Larcher, K.~A. Lee, B.~Ma, and H.~Li,
\newblock ``{T}ext-dependent {S}peaker {V}erification: {C}lassifiers,
  {D}atabases and {RSR2015},''
\newblock {\em Speech Communucation}, vol. 60, pp. 56--77, 2014.

\bibitem{chung2020generative}
Y.-A. Chung and J.~Glass,
\newblock ``{G}enerative pre-training for speech with auto-regressive
  predictive coding,''
\newblock in {\em Proc of ICASSP}, 2020, pp. 3497--3501.

\bibitem{tensorflow2015-whitepaper}
M.~Abadi et~al.,
\newblock ``{TensorFlow}: large-scale machine learning on heterogeneous
  systems,'' 2015,
\newblock Software available from tensorflow.org.

\bibitem{Povey_ASRU2011}
D.~Povey et~al.,
\newblock ``{T}he kaldi speech recognition toolkit,''
\newblock in {\em Proc. on Automatic Speech Recognition and Understanding
  (ASRU)}, 2011.

\bibitem{SRE08}
``https://www.nist.gov/itl/iad/mig/2008-nist-speaker-recognition-evaluation-results,''
  .

\end{thebibliography}
\end{document}